\documentstyle[psfig,aps]{revtex}

\newcommand{\bms}[1]{\mbox{\boldmath$#1$}}
\newcommand{\bmss}[1]{\mbox{\boldmath$\scriptstyle#1$}}
\newcommand{\tfrac}[2]{\mbox{$\frac{#1}{#2}$}}

\newcommand{\ket}[1]{| #1 \rangle}

\begin{document}
\twocolumn
\title{
Experimental Quantum Error Correction
}
\draft
\author{
D.~G. Cory$^1$, W. Mass$^1$, M. Price$^1$
E. Knill$^2$, R. Laflamme$^2$ W.~H. Zurek$^2$
T.~F. Havel$^3$ and S.~S. Somaroo$^3$
}
\address{
$^1$Dept.\ of Nuclear Engineering,
Massachusetts Institute of Technology,
Cambridge, MA 02139 \\
$^2$Los Alamos National Laboratory,Los Alamos, NM 87545 \\
$^3$BCMP, Harvard Medical School,
240 Longwood Ave., Boston, MA 02115
}
\date{\today}
\maketitle
\begin{abstract}
Quantum error correction is required to compensate for the fragility
of the state of a quantum computer.  We report the first experimental
implementations of quantum error correction and confirm the expected
state stabilization.  In NMR computing, however, a net improvement in
the signal-to-noise would require very high polarization.  The
experiment implemented the 3-bit code for phase errors in liquid state
state NMR.
\end{abstract}

Quantum computers exploit the superposition principle
to solve some problems much more efficiently
than any known algorithm for their classical counterparts.
These problems include factoring large numbers\cite{shor:qc1994a}
(thereby breaking public key cryptography), combinatorial
searching\cite{grover:qc1995a} and simulations of quantum
systems\cite{feynman:qc1986a,lloyd:qc1996a,zalka:qc1996b}.
Exploiting the power of quantum computation was thought to be
physically impossible due to the extreme fragility of quantum
information\cite{landauer:qc1995a,unruh:qc1995a}.
This judgment was demonstrated to be overly pessimistic when quantum error-correction
techniques\cite{shor:qc1995b,steane:qc1995a,shor:qc1996a}
were found to protect quantum information against
corruption due to imperfect control and decoherence. 
It is now known that for physically reasonable models
of decoherence a quantum computation can be as long
as desired with arbitrarily accurate answers,
provided the error rate is below a threshold
value\cite{knill:qc1998a,aharonov:qc1996a,preskill:qc1998a,kitaev:qc1996a}.
Thus decoherence and imprecision are no longer considered
insurmountable obstacles to realizing a quantum computer.

The chief remaining obstacle to quantum computing is the
difficulty of finding suitable physical systems whose quantum states
can be accurately controlled.
Devices based on ion traps \cite{cirac:qc1995a} have
so far been limited to two bits\cite{monroe:qc1995a}.
Recently, liquid state NMR techniques have been shown
to be capable of quantum computations with at least
three bits~\cite{cory:qc1997b,laflamme:qc1997a}.
This makes it possible, for the first time,
to experimentally implement the simplest quantum
error-correcting codes, and so test these ideas in physical
systems with a variety of decoherence, dephasing and relaxation phenomena.

In room temperature liquid state NMR, one can coherently manipulate
the internal states of the coupled spin $1\over 2$ nuclei in each of
an ensemble of molecules subject to a large external magnetic field.
Although the ensemble nature of the system and the small energy
of each spin imply that the set of accessible states is highly mixed,
it has been shown that experimental methods exist that can be used to
isolate the pure state behavior of the system, thus permitting limited
application of NMR to quantum computation\cite{cory:qc1997a,chuang:qc1997a}.

Here we describe the implementation of a quantum
error-correcting code which compensates for small phase errors due to
fluctuations in the local magnetic field. The behavior of this code
was measured for two systems: The ${}^{13}$C labeled carbons in
alanine subject to the correlated phase errors induced by diffusion in
a pulsed magnetic field gradient, and the proton and two labeled carbons
in trichloroethylene (TCE) subject to its natural relaxation
processes.  In alanine, we observed correction of first-order errors
for a given input state, while in TCE we demonstrated
state preservation of an arbitrary state
by error correction.

In NMR loss of phase coherence may be both due to bona fide
irreversible decoherence\cite{zurek:qc1991} and due to dephasing which can be reversed by
spin echo.  Both lead to symptoms which, in an ensemble, are identical from the
standpoint of quantum computation. Both can be corrected using the
same scheme we shall demonstrate below.

It is important to comment on the applicability
of this technique to ensemble quantum computing.
Although our experiments validate the usefulness of
error correction for quantum computing with pure states,
there is a substantial loss of signal associated with
the use of ancilla spins in weakly polarized systems.
We argue that in this setting,
the loss of signal involved in exploiting ancillas
removes any advantage for computation gained by error correction,
at least unless the system is sufficiently polarized
to enable the generation of nearly pure states.
Nevertheless, our experiments demonstrate that error-correcting
codes can be implemented, and that they behave as predicted.


The simple three-bit quantum error-correcting
code used here is designed to compensate to
first order for small random phase fluctuations.
These fluctuations constitute a random evolution of the state
\begin{eqnarray} \label{eq:phase}
\ket{b_1b_2b_3} & ~\longrightarrow~ &
   e^{-i(\theta_1\bmss\sigma_{\rm z}^{1}+\theta_2\bmss\sigma_{\rm
   z}^{2}+\theta_3\bmss\sigma_{\rm z}^{3})}
   \ket{b_1b_2b_3}
\\ && =~ \nonumber
   e^{i( (-1)^{b_1}\theta_1+(-1)^{b_2}\theta_2+(-1)^{b_3}\theta_3)}
   \ket{b_1b_2b_3} ~,
\end{eqnarray}
where $b_i$ is $0$ or $1$, $\theta_i$ is a random phase variable,
and $\bms\sigma_z^{i}$ is the Pauli matrix acting on the $i$'th spin.
The $\theta_i$ depend on the error rates in the
model, which is described in detail below.

The error-correcting code itself is a phase variant
of the classical three bit majority code with a decoding technique
that preserves the quantum information in the encoded state.
Let $\ket{+} = (\ket{0}+\ket{1})/\sqrt{2}$
and $\ket{-} = (\ket{0}-\ket{1})/\sqrt{2}$.
The state $(\alpha\ket{000}+\beta\ket{100})$ is encoded as
$\alpha\ket{+\,+\,+}+\beta\ket{-\,-\,-}$ by a unitary transformation.
The first-order expansion of the operator in Eq.\ (\ref{eq:phase})
in the small random phases is
\begin{equation}
{\bf 1} - i\theta_1\bms\sigma_z^{1} - i\theta_2\bms\sigma_z^{2}
- i\theta_3\bms\sigma_z^{3} ~,
\end{equation}
which evolves the encoded state to 
\begin{eqnarray}
\alpha\ket{++\,+}+\beta\ket{--\,-}&\rightarrow& ~~
\alpha\ket{++\,+}+\beta\ket{--\,-}\\ \nonumber && - i \theta_1
(\alpha\ket{-+\,+}+\beta\ket{+-\,-})\\ \nonumber && - i \theta_2
(\alpha\ket{+-\,+}+\beta\ket{-+\,-})\\ \nonumber && - i \theta_3
(\alpha\ket{++\,-}+\beta\ket{--\,+}) ~. \nonumber
\end{eqnarray}

\begin{figure}
\begin{center}
\mbox{
\psfig{figure=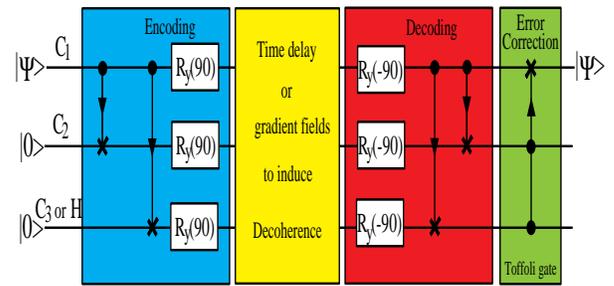,height=1.5in,width=3.2in}
}
\end{center}
\caption{Network for encoding, decoding and error correction.
The circuit describes the evolution of the 3 bits as a function of time.
The gate $\bullet$\hskip -0.05truein ---\hskip -0.1truein $>$\hskip -0.02truein --\hskip -0.05truein {\tt x}
 corresponds to a control-not, i.e a gate which flips
the target bit ({\tt x}) if the control bit ($\bullet$) is in the state $|1\rangle$.
$R_y(90)$ represent  a single bit gate  corresponding
to a rotation by an angle of $\pi/2$
around the $y$-axis.  The Tofolli gate flips the target bit ({\tt x})
if the two control bits ($\bullet$) are in the state $|1\rangle$.
A detailed implementation of these gates is given in [17].
The information carrying bit is carbon 1 (see Figures 2 and 3)
in both experiments.
}
\label{figure:dec_net}
\end{figure}

The different errors map the encoded state into orthogonal subspaces. 
Thus one can determine which error occurred
without destroying the encoded quantum information. This is
accomplished by a measurement which reveals the subspace the
state has moved into.  After decoding, the original state of the first
spin can then be restored by a unitary transformation, while the other
two spins contain
information (the syndrome) about the error which occurred.  A network
which accomplishes the encoding, decoding and error-correction steps
is shown in Figure~\ref{figure:dec_net}.


In NMR experiments, non-unitary processes
are classified as spin-lattice and
spin-spin relaxation~\cite{OldTestiment,NewTestiment}.
The former involves an exchange of energy
with the non-spin degrees of freedom (the lattice),
and returns the spins to thermal equilibrium.
The latter leads to a loss of phase coherence.
For spin $\frac{1}{2}$ nuclei, both processes
are due to fluctuating local magnetic
fields. The three spin code corrects
for errors due to locally fluctuating fields
along the z axis.

We focus on a weakly coupled three-spin system where the
strongest contribution to coherence loss is
from external fields which contribute the Hamiltonian
\begin{equation}
{\cal H}_{\rm R}~\equiv~\gamma^1 {\bf I}^1 \cdot {\bf B}^1(t) +
\gamma^2 {\bf I}^2 \cdot {\bf B}^2(t) +
\gamma^2 {\bf I}^3 \cdot {\bf B}^3(t)~,
\end{equation}
where ${\bf I} = ({\bf I}_{\rm x},{\bf I}_{\rm y},{\bf I}_{\rm z})$
and ${\bf I}_u = {1\over 2}\bms\sigma_u$ ($u={\rm x,y,z}$).  In the
case of slowly varying external fields the induced random phase
fluctuations are identical to those described in Eq.~(\ref{eq:phase}).
As a result, the off-diagonal elements of the density matrix decay
exponentially at a rate which depends on the fields ${\bf B}^k$ at
each spin, their gyromagnetic ratios $\gamma^k$, the coherence order
and the zero frequency components of the spectral densities of the
fields.
The ``coherence order'' is the difference
between the total angular momenta along the z-axis of the two states
$\ket{b}$, $\ket{b'}$ (in units of $\hbar/2$)\cite{gradientRef}.  

To obtain a clean demonstration of error correction,
a simple error model was implemented precisely in the case of alanine.
This implementation used the random molecular motion
induced by diffusion in a constant field gradient
to mimic the effect of a slowly varying random field.
This is achieved by turning on an external field gradient
$\nabla_{\rm z} {\bf B} = \partial B_{\rm z} / \partial z$ across
the sample for a time $\delta$.
This modifies the magnetization in the sample
with a phase varying linearly along the z direction according to
$\partial\phi/\partial z = n \delta \gamma \partial B_{\rm z}/\partial z$,
where $n$ is the coherence order of the density matrix element and
$\gamma$ is the gyromagnetic ratio.
A reverse gradient is used to refocus
the magnetization after allowing molecular diffusion
to take place for amount of time $t$.
As a result of random spin displacement $\Delta z$,
the phases of the spins
are not returned to their original values but are randomly modified by
$(n\delta\gamma\partial B_{\rm z}/\partial z)\Delta z$.
For a gaussian displacement profile with a width
of $\sqrt{2Dt}$,
the effective decoherence time of this process
is proportional to the diffusion constant $D$
as well as to the square of the coherence order $n$~\cite{gradientRef}:
\begin{equation}
\frac{1}{\tau} 
~=~ \left( \frac{\partial\phi}{\partial z}\right)^2 D
~=~\gamma^2 (\nabla_{\rm z} {\bf B})^2 n^2 \delta^2 D ~.
\end{equation}
This artificially induced ``decoherence'' in the alanine experiments is an
example of completely correlated phase scrambling. This occurs
naturally if all the spins have equal gyromagnetic ratios in the slow
motion regime.
We used TCE to demonstrate error-correction in the
presence of the natural decoherence and dephasing of a molecule where
$T_2\ll T_1$.


Most NMR experiments are described
using the product operator formalism \cite{SoEiLeBoEr:83}.
This formalism  describes the state as a sum of products
of the operators ${\bf I}_x^k$, ${\bf I}_y^k$, ${\bf I}_z^k$.
The identity component of such a sum is the same
for any state and is usually suppressed to yield
the ``deviation'' (traceless) density matrix.
The effect of error-correction can be understood from
the point of view of this formalism. As an example,
consider encoding the
state ${\bf I}_{\rm z}^1$ using two ancillas
initially in their ground states.
Up to a constant factor, the initial state is described by
\begin{eqnarray} 
\mbox{\boldmath$\rho$}_{\sf A} &=&
{\bf I}_{\rm z}^1
(\tfrac{1}{2} {\bf 1} + {\bf I}_{\rm z}^2)
(\tfrac{1}{2} {\bf 1} + {\bf I}_{\rm z}^3)\nonumber\\
&=& \tfrac{1}{4} {\bf I}_{\rm z}^1 +
\tfrac{1}{2} {\bf I}_{\rm z}^1 {\bf I}_{\rm z}^2 +
\tfrac{1}{2} {\bf I}_{\rm z}^1 {\bf I}_{\rm z}^3 +
{\bf I}_{\rm z}^1 {\bf I}_{\rm}^2 {\bf I}_{\rm z}^3 ~.
\label{eq:input_states}\label{eq:rhoA}
\end{eqnarray}
The density matrix $\bms\rho_{\sf A}$
consists of an incoherent sum of four terms.

After encoding the state is
\begin{equation}
\bms\rho_{\mathsf B} ~\equiv~ \tfrac{1}{4} \left(
{\bf I}_{\mathrm x}^1 + {\bf I}_{\mathrm x}^2 +
{\bf I}_{\mathrm x}^3 + 4\, {\bf I}_{\mathrm x}^1
{\bf I}_{\mathrm x}^2 {\bf I}_{\mathrm x}^3 \right) ~.
\end{equation}
In the case of completely correlated phase errors,
this decays as
\begin{eqnarray}
\bms\rho_{\sf C} &~\equiv~& \tfrac{1}{4} \left(
\left( {\bf I}_{\rm x}^1 + {\bf I}_{\rm x}^2 +
{\bf I}_{\rm x}^3 \right) e^{-t / \tau} \right.\\&&\left.\mbox{}\hspace{3.5pt} +
\left( 3 {\bf I}_{\rm x}^1 {\bf I}_{\rm x}^2 {\bf I}_{\rm x}^3
+ {\bf I}_{\rm x}^1 {\bf I}_{\rm y}^2 {\bf I}_{\rm y}^3
+ {\bf I}_{\rm y}^1 {\bf I}_{\rm x}^2 {\bf I}_{\rm y}^3
+ {\bf I}_{\rm y}^1 {\bf I}_{\rm y}^2 {\bf I}_{\rm x}^3 \right)
e^{-t / \tau} \right. \nonumber\\ && \left. \mbox{}\hspace{3.5pt} +
\left( {\bf I}_{\rm x}^1 {\bf I}_{\rm x}^2 {\bf I}_{\rm x}^3
- {\bf I}_{\rm x}^1 {\bf I}_{\rm y}^2 {\bf I}_{\rm y}^3
- {\bf I}_{\rm y}^1 {\bf I}_{\rm x}^2 {\bf I}_{\rm y}^3
- {\bf I}_{\rm y}^1 {\bf I}_{\rm y}^2 {\bf I}_{\rm x}^3 \right)
e^{-9 t / \tau} \right) ~. \nonumber
\end{eqnarray}
Decoding and error correction
mixes these states together so as to cancel
the initial decay of the first spin.
The reduced
density matrix for the first spin becomes
\begin{equation} \label{eq:rhoE}
\bms\rho_{\sf E}^1 ~\equiv~ 
\tfrac{1}{8} {\bf I}_{\mathrm z}^1
( 9 e^{-t/\tau} - e^{-9t/\tau} ) ~\approx~
 {\bf I}_{\mathrm z}^1
( 1 - {9\over 2} t^2 / \tau^2 + \cdots ) ~.
\end{equation}
The effect of error correction can be seen from the absence of terms
depending linearly on $t$.  Similar results can be obtained for other
initial states and also for uncorrelated phase errors.



\begin{figure}
\begin{center}
\mbox{
\psfig{figure=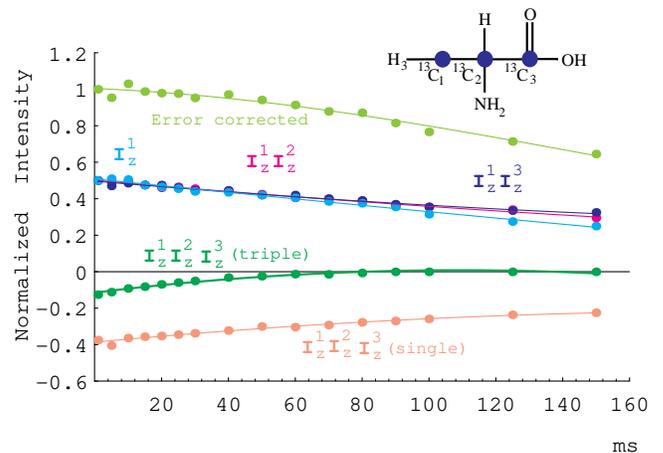,height=2.5in,width=3.2in}
}
\end{center}
\caption{The intensities of the magnetization of the first
spin after applying the dephasing and decoding
procedures described in the text,
together with single exponential fits to the
intensities versus the dephasing time $\tau$.
The three mixed states ${\bf I}_{\rm z}^1$,
${\bf I}_{\rm z}^1 {\bf I}_{\rm z}^2$,
${\bf I}_{\rm z}^1 {\bf I}_{\rm z}^3$,
evolved as single quantum coherences during
$\tau$, whereas
${\bf I}_{\mathrm z}^1 {\bf I}_{\mathrm z}^2 {\bf I}_{\mathrm z}^3$
evolved as a mixture of single and triple quantum coherences,
which have been plotted separately (single and triple).
The sum of these intensities and the corresponding fits
(Error corrected) give the intensities and fit characteristics of
the same experiment using a pseudopure state (see text).
The initial slope of the total is close to zero, thus showing
that the error-correction procedure was able to cancel
loss of coherence of the \protect\linebreak pseudopure state to first order.
} \label{figure:ala_res}
\end{figure}

In the alanine experiments, each of the
four product operators in the sum of Eq.~\ref{eq:input_states}
was realized
in a separate experiment, and the final state after
encoding and decoding inferred by adding the results.
The loss of polarization over time in each product operator was
measured explicitly in each experiment. The results were added
computationally to simulate the effect of the Toffoli gate and are
shown in Figure~\ref{figure:ala_res}. The components of ${\bf
I}_{\mathrm z}^1 {\bf I}_{\mathrm z}^2 {\bf I}_{\mathrm z}^3$ which
evolved as a single and a triple coherence when encoded were separated
by gradient labeling~\cite{gradientRef}.  
The initial slopes of the decay curves for each
operator were estimated by linear square fit to the logarithm of the
measured intensities. The values obtained are $-0.0034$ ( ${\bf
I}_{\mathrm z}^1$), $-0.0030$ ($2 {\bf I}_{\mathrm z}^1 {\bf
I}_{\mathrm z}^2$), $-0.0047$ ($2 {\bf I}_{\mathrm z}^1 {\bf
I}_{\mathrm z}^3$), $-0.0038$ ($4 {\bf I}_{\mathrm z}^1 {\bf
I}_{\mathrm z}^2 {\bf I}_{\mathrm z}^3$, single coherence) and
$-0.0395$ ($4 {\bf I}_{\mathrm z}^1 {\bf I}_{\mathrm z}^2 {\bf
I}_{\mathrm z}^3$, triple coherence) When the slopes are added as
required for error-correction with a pseudopure input, we get a
measured net slope of $0.00081$, which should be compared to the
slopes for the single coherences whose average is $-0.0037$. The net
curve has quadratic behavior for small delays to within experimental
error.

\begin{figure}
\begin{center}
\mbox{
\psfig{figure=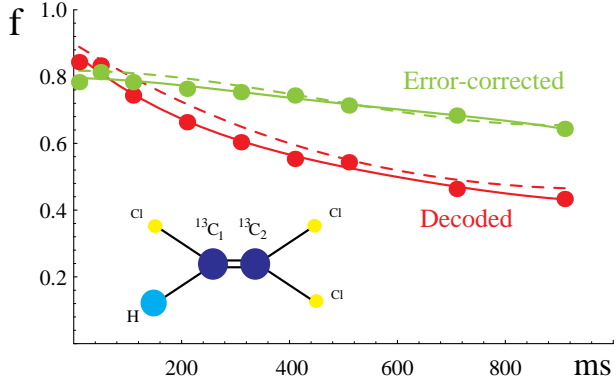,width=3in,height=2in}
}
\end{center}
\caption{Experimentally determined entanglement fidelities for the TCE
experiments after decoding (red) and after decoding and error
correction (green). The continuous curves are interpolations of the
data points. The broken curves were determined by simulating the pulse
sequence using the measured coupling constants and estimated $T_2$'s
of $1.1s$ (C1), $0.6s$ (C2) and $3s$ (H). Differences between
experimental and theoretical curves are attributed to lack of precise
knowledge of the error model. Errors in the data points are
approximately $0.05$. Note that since the proton $T_2$ is much longer
than that of the carbons, the long term gain in fidelity is partially
due to recovery of polarization from the proton. The demonstration of
error correction lies in the initial slope. The curves show that error
correction decreases the initial slope by a factor of $\sim 10$ (by
square fit to the logarithm).}
\label{figure:tce_fids}
\end{figure}

The goal of our experiments with TCE was to establish
the behavior of encoding/decoding and error correction on
all possible initial states subject to the natural decoherence and dephasing.
The spins were prepared in
the states 
\begin{equation}
\rho({1\over 2}{\bf 1}+{\bf I}_z^{2})({1\over 2}{\bf 1}+{\bf
I}_z^{3}),
\label{eq:tceinput}
\end{equation}
with $\rho$ one of the four inputs ${1\over 2}{\bf 1}$, ${\bf I}_z^1$,
${\bf I}_x^1$, ${\bf I}_y^1$. Any possible input is just a linear
combination of these four states.  We used gradient methods to
directly generate the four states of Eq.~\ref{eq:tceinput} with a
series of experiments.  They were then subjected to pulse sequences
for encoding and decoding (experiment I) with a variable delay between
the two operations. Decoherence and dephasing take place during the delay. The
reduced density matrix on the first spin (the output) was measured.
In the second experiment (II) decoding was followed by error
correction before the output was determined.  Ideally the output would
be identical to the input. The measured outputs were compared to the
ideal ones by computing the ``entanglement
fidelity''~\cite{entanglementfid}.  This is a useful measure of how
well the quantum information in the input is preserved.  Entanglement
fidelity is the sum of the correct polarization left in the output
state for each input. More precisely, given input ${\bf I}_a^1$, let
$f_a$ be the relative polarization of ${\bf I}_a^1$ in the output
compared to the input.  Then $f={1\over 4}(1+f_x+f_y+f_z)$, this
formula is correct for processes which do not affect the completely
mixed state ${1\over 2}{\bf 1}$.  The results for nine different
delays are shown in Figure~\ref{figure:tce_fids}.  The curves show
that error correction decreases the initial slope by a factor
of $\sim 10$ (by square fit to the logarithm).


Our demonstration of error-correction does not
imply that error-correction can be used to overcome the problems of
high temperature ensemble quantum computing.  In this model of quantum
computing, the initial state can be described as a small, linear
deviation from the infinite temperature equilibrium. Thus, the
deviation is proportional to a Hamiltonian of $n$ weakly interacting
particles. In this limit no method of error-correction based on
externally applied, time dependent fields can improve the polarization
of any particle by more than a factor proportional to $\sqrt{n}$
\cite{sorensen:qc1989}. If one wishes to use error-correction 
an even bigger problem is encountered:
The initial state of the ancillas used
for each encoding/decoding cycle must be pure.  In the high
temperature regime, the best we can do to ensure that is to generate
a pseudopure deviation in the ancillas.  Unfortunately, this
deviation has to be created {\em simultaneously\/} on all ancillas,
leading to an exponential reduction in polarization proportional to
the total number of ancillas required~\cite{warren:qc1997a}.  This
always exceeds the loss of polarization.  
Another problem is the inability to reuse ancilla bits.
This has two consequences. The
first is that decoherence rapidly removes information in the state,
leading to computations which are logarithmically bounded in
time~\cite{aharonov:qc1996b}. Second, the total number of ancillas
required is proportional to the time-space product of the computation,
rather than to a power of it's logarithm.

Our work shows that liquid state NMR can be used to test fundamental
ideas in quantum computing and communication involving non-trivial numbers
of bits subject to a variety of errors.
Our experiments demonstrate for the first time the state preserving
effect of the three bit phase error-correcting code. 
The first-order behavior was established to high
accuracy for a specific state in alanine, while the overall effect
was observed and the improvement in state recovery verified
in TCE. These experiments confirm not only the validity of theories of
quantum error correction in a simple case, but also demonstrate the
ability, in liquid state NMR, to control the state of three spin-half
particles.  This is an important advance for quantum computing, as
this is the first system where this degree of control has been
successfully implemented.

\bigskip\centerline{\bf Acknowledgments}
We thank Jeff Gore for his help with the simulations.
This work was supported by, or in part by,
the U.\ S.\ Army Research Office under
contract/grant number DAAG 55-97-1-0342
from the DARPA Ultrascale Computing Program. E.~K., R.~L. and W.~H.~Z
thank the National Security Agency for support.

\bibliographystyle{plain}

\vfill\pagebreak


\vfill
\mbox{}


\end{document}